\documentclass[10pt,conference]{IEEEtran}
\IEEEoverridecommandlockouts
\usepackage{cite}
\usepackage{amsmath,amssymb,amsfonts}
\usepackage{algorithmic}
\usepackage{graphicx}
\usepackage{textcomp}
\usepackage{xcolor}
\usepackage{todonotes}
\usepackage{url}
\usepackage{graphicx}
\usepackage{array}
\usepackage{menukeys}
\usepackage{svg}
\usepackage[framemethod=tikz]{mdframed}
\usepackage{lipsum}

\newcommand{\etal}{\emph{et al.}\ }
\def\BibTeX{{\rm B\kern-.05em{\sc i\kern-.025em b}\kern-.08em
    T\kern-.1667em\lower.7ex\hbox{E}\kern-.125emX}}

\usepackage[most]{tcolorbox}
\tcbset {
  base/.style={
    arc=0mm, 
    bottomtitle=0.5mm,
    boxrule=0mm,
    colbacktitle=black!10!white, 
    coltitle=black, 
    fonttitle=\bfseries, 
    left=2.5mm,
    leftrule=1mm,
    right=3.5mm,
    title={#1},
    toptitle=0.75mm, 
  }
}

\definecolor{brandblue}{rgb}{0.34, 0.7, 1}
\newtcolorbox{mainbox}[1]{
  colframe=brandblue, 
  base={#1}
}

\newtcolorbox{subbox}[1]{
  colframe=black!30!white,
  base={#1}
}

\definecolor{antiquewhite}{rgb}{0.98, 0.92, 0.84}
\definecolor{arsenic}{rgb}{0.23, 0.27, 0.29}

\newmdenv[innerlinewidth=0.5pt, roundcorner=4pt,linecolor=arsenic,backgroundcolor=antiquewhite,innerleftmargin=6pt,
innerrightmargin=6pt,innertopmargin=6pt,innerbottommargin=6pt]{promptbox}

\begin{document}

\title{An Exploratory Study of ML Sketches and Visual Code Assistants}

\makeatletter
\newcommand{\linebreakand}{%
  \end{@IEEEauthorhalign} 
  \hfill\mbox{}\par       
  \mbox{}\hfill           
  \begin{@IEEEauthorhalign} 
}
\makeatother

\author{
\IEEEauthorblockN{Luís F. Gomes}
\IEEEauthorblockA{\textit{Software and Societal Systems Dept.} \\
\textit{Carnegie Mellon University}\\
Pittsburgh, USA  \\
lfgomes@andrew.cmu.edu}

\and
\IEEEauthorblockN{Vincent J. Hellendoorn}
\IEEEauthorblockA{\textit{Software and Societal Systems Dept.} \\
\textit{Carnegie Mellon University}\\
Pittsburgh, USA \\
vhellendoorn@cmu.edu}

\linebreakand 

\IEEEauthorblockN{Jonathan Aldrich}
\IEEEauthorblockA{\textit{Software and Societal Systems Dept.} \\
\textit{Carnegie Mellon University}\\
Pittsburgh, USA \\
jonathan.aldrich@cs.cmu.edu }

\and 

\IEEEauthorblockN{Rui Abreu}
\IEEEauthorblockA{\textit{INESC-ID} \\
\textit{Faculty of Engineering, University of Porto}\\
Porto, Portugal \\
rui@computer.org}

}
\maketitle

\begin{abstract}
This paper explores the integration of Visual Code Assistants in Integrated Development Environments (IDEs). In Software Engineering, whiteboard sketching is often the initial step before coding, serving as a crucial collaboration tool for developers. Previous studies have investigated patterns in SE sketches and how they are used in practice, yet methods for directly using these sketches for code generation remain limited. The emergence of visually-equipped large language models presents an opportunity to bridge this gap, which is the focus of our research.
In this paper, we built a first prototype of a Visual Code Assistant to get user feedback regarding in-IDE sketch-to-code tools.
We conduct an experiment with 19 data scientists, most of whom regularly sketch as part of their job.
We investigate developers' mental models by analyzing patterns commonly observed in their sketches when developing an ML workflow. Analysis indicates that diagrams were the preferred organizational component (52.6\%), often accompanied by lists (42.1\%) and numbered points (36.8\%).
Our tool converts their sketches into a Python notebook by querying an LLM. We use an LLM-as-judge setup to score the quality of the generated code, finding that even brief sketching can effectively generate useful code outlines.
We also find a positive correlation between sketch time and the quality of the generated code.
We conclude the study by conducting extensive interviews to assess the tool's usefulness, explore potential use cases, and understand developers' needs. As noted by participants, promising applications for these assistants include education, prototyping, and collaborative settings.
Our findings signal promise for the next generation of Code Assistants to integrate visual information, both to improve code generation and to better leverage developers' existing sketching practices.

\end{abstract}


\section{Introduction}

\subsection{Motivation}

Developing software is a collaborative process that relies on the collective creativity and problem-solving skills of a team \cite{williams2000collaborative}. It often begins with brainstorming sessions where developers gather around a whiteboard to sketch out initial ideas and concepts \cite{software_mangano_2010}. This collaborative sketching helps in aligning everyone's understanding, creating a more solid foundation for the subsequent stages of development. Developers also use sketches as a form of planning their own code \cite{go_cherubini_2007, visual_walny_2011}, communicating and expressing their mental models through sketches, as in the typical whiteboard meetings \cite{sketches_baltes_2017}. Drawings and diagrams thus serve as visual aids to express and refine mental models of the code structure \cite{maintaining_latoza_2006}.

This ties into a recent development in Artifical Intelligence research: Large Language Models (LLMs) have evolved towards multi-modality, which includes strong vision capabilities \cite{geminiteam2024gemini, gpt4_openai_2023}. 
While the use of natural language to generate code (text-to-code) is already widespread and a common focus of research with many Code LLMs being developed \cite{chen2021evaluating, guo2024deepseekcoderlargelanguagemodel, roziere2024codellama}, using image-to-code techniques has almost exclusively been explored for tasks such as front-end development \cite{si2024automatingfrontend} where the sketch directly captures the visual layout that the code should provide. This is orthogonal to sketches that capture the structure of a software system.

In this work, we bridge the gap between existing research on whiteboard sketches and their use to generate code, focusing specifically on Data Science (DS) systems. The demand for DS software has recently increased steeply \cite{Krenn_2023}. Such systems typically involve a pipeline that transforms data in various ways, generating models and other artifacts (e.g., visualizations) along the way. As shown in this study, data scientists frequently sketch their systems and such sketches tend to be fairly rich in detail. More generally, our work explores the concept of \textit{Visual Code Assistants}. Complementing the typical text-based assistants, a Visual Code Assistant is capable of understanding developers' sketches, helping them when they express their ideas visually.

\subsection{Contributions}

This paper makes contributions to both the research community and developers by offering guidance on developing generative code tools and providing insights into the mental models of ML developers. We implement a visual assistant by building an in-IDE sketch-to-code tool, which converts drawings of DS pipelines into IPython Jupyter Notebooks. We conducted a user study with 19 data scientists, the majority of whom regularly sketch. The study is composed of two parts: \textbf{sketching} and \textbf{programming}. In the first part, a task is given to participants and they are asked to behave as in a regular team meeting (or code interview), using the whiteboard to explain how they would solve it. In the second part, we use the Visual Code Assistant to generate an initial code skeleton and the participant iteratively edits it to achieve the final solution. Both parts follow the think-aloud protocol \cite{Ericsson1993}. The study is guided by three research questions:

\begin{itemize}
    \item \textbf{RQ1}: What are the prevalent sketching patterns of ML developers?
    \item \textbf{RQ2}: To what extent can current sketch-to-code technologies support ML developers?
    \item \textbf{RQ3}: What are developers' perspectives on the value and potential of in-IDE Visual Code Assistants?
\end{itemize}

\noindent
We observe that real drawings are highly varied, frequently mixing diagrams and textual content. Using a separate LLM to score the generated code against a set of requirements, we find that popular LLMs can generate valid outlines (the high-level structure) of the desired notebooks with 70\%-80\% accuracy, and implement 25\%-40\% of the requirements correctly from sketches alone. Longer sketch times correlate positively with outline quality, but not (significantly) with implementation details, in part because sketches often do not provide all details required to implement the code correctly. We collect rich qualitative insights from interviews conducted after the study. The contributions of our work are as follows:

\begin{enumerate}
\item We develop an in-IDE Visual Code Assistant prototype that converts sketches of ML-related tasks into complete Jupyter Notebooks.
\item We conduct the first exploratory study on \textit{sketching} for ML-related tasks and the integration of vision capabilities in Code Assistants.
\item We carry out the first user study to evaluate the quality of code generated by LLMs based solely on images.
\item We provided guidance for future Visual Code Assistant development by reporting on developers' needs and identifying potential use cases where this technology can be most effectively applied.
\end{enumerate}

\section{Related Work}

Programming is all about understanding tasks and creating code to execute them. To effectively work on software development, programmers must create a clear understanding of both the current code structure and any new code they need to write. This involves constructing mental models that help them visualize how the code works and how different components interact with each other.

\subsection{Sketching in SE}

In a survey with 280 software engineers, LaToza \etal \cite{maintaining_latoza_2006} studied this phenomenon in 2006, arguing how mental models of code are expensive to create and maintain. Yet, they found that these mental models are only kept in developers' heads or ``in transient forms such as sketches on a whiteboard''. Furthermore, they pointed out how ``paper and whiteboards were perceived most effective'' while designing software.
They suggested reducing the cost of using design documents by linking them to the code or building ``tools that capture informal whiteboard or paper designs''.

Building on LaToza \etal's findings, many researchers studied sketching practices in SE. Cherubini \etal
showed that developers often express their code in temporary drawings that are later lost, due to the cost of translating them into electronic renderings. The transient nature of whiteboard sketches can be reduced by Visual Coding Assistants. When uploading their sketches to generate code, developers make them persistent as a side effect. Branham \etal 
built ReBoard, which captures whiteboard drawings. They studied how whiteboard content is reused, identifying issues to be addressed when building augmentation systems. Similarly, Mangano \etal \cite{software_mangano_2010 } identified three basic needs of developers when sketching: frequently shifting focus, using low detailed models, and using a mix of notations. They presented positive results with Calico, a tool to address these issues. We find that the flexibility of current AI vision models makes them suitable for working with this mix of notations, low-detailed models, and informal sketches. As developers frequently shift their focus, these models should be integrated into the IDE.

Walny \etal \cite{follow_walny_2011} followed the lifecycle of sketches and diagrams, reporting on their transitions and commenting on how informal sketches can usefully serve as memory aids and for communication purposes. In a second study \cite{visual_walny_2011}, they observed 82 whiteboard sketches, concluding that participants have an ``inventive capability to create [...] representation of whatever problems, processes or data they wanted to discuss''. They also found that words were often used as primary objects and not only as labels. This highlights that developers have the ability to sketch different problems, and drawing is a natural way of expression.

In ``Sketches and Diagrams in Practice'' \cite{sketches_baltes_2017}, Baltes \etal indicated that the majority of sketches were created informally on analog media, and were found to be valuable resources for understanding source code artifacts. They also pointed out that sketches usually represent higher levels of abstraction. 
They propose two new tools to bridge analog and digital artifacts: SketchLink \cite{linking_baltes_2017} links code to sketches using a web app to capture the drawings and an IDE plugin to visualize them, and LivelySketches  \cite{roundtrip_baltes_2017} supports the iterative process of sketching, versioning and linking drawings. More recently, Almeida \etal \cite{go_almeida_2022} interviewed 27 software architects, followed by a survey with 46 participants. Their findings contradict prior research, affirming that 76\% of software architects actually document their whiteboard meetings, with photos of the whiteboard among the five most common documentation approaches. However, they also observe that in whiteboard meetings there is not ``sufficient information about the problem to design the solution'' and that ``certain aspects of the solution are over-simplified'' which makes the solution mismatch the original sketch. This matches the rate and characteristics of sketching in our work. These studies suggest that visual assistants should expect to mainly extract ``big picture" information from sketches, rather than low-level implementation details.

\subsection{Images to Code}

The use of sketches as a specification for code generation has been explored previously, especially when the code is highly related to visuals. This is the case in graphic interface design, where the use of hand-drawn mock-ups is common and its translation to HTML code is very useful. Pix2Code \cite{beltramelli2017pix2code} explored this problem and uses Convolutional and Recurrent Neural Networks to generate code from designers’ mock-ups. Ellis \etal \cite{learning_ellis_2018} convert drawings into LaTeX programs representing the intended graphics. Although not generating code, PSDoodle \cite{psdoodle_mohian_2022} uses sketches as the input to perform an app screen search. On ML code generation, Sethi \etal present DLPaper2Code \cite{dlpaper2code_sethi_2017} that generates Python code from images of Deep Learning architectures.

With the advent of multi-modal LLMs, a range of innovative use cases are being successfully explored. Notable contributions include MathVista \cite{mathvista24} and MMMU \cite{yue2024mmmu}, which investigate the integration of multi-modal capabilities in mathematics and college-level reasoning, respectively. In the domain of frontend development, Si \etal \cite{si2024automatingfrontend} examine the performance of GPT-4 and Gemini in generating accurate web pages from images. Additionally, Pădurean and Singla \cite{pădurean2024} analyze visual programming generation performance tailored for elementary-level students. Plot2Code \cite{wu2024plot2code} extends the application of LLMs to code generation from scientific plots, while MMCode \cite{li2024mmcode} focuses on multi-modal coding for evaluating algorithmic problem-solving skills in visually rich contexts. D'Amorim \etal \cite{visual_damorim_2020}, proposed the concept of sketch-to-code applied to the domain of DS, arguing that it is a highly visual discipline.
We build on their vision, aiming for the broader goal of operating on complete data workflows, and generating Jupyter Notebooks based on whiteboard drawings.

In this paper, we introduce the first benchmark specifically designed to assess the performance of leading LLMs in interpreting and generating code from whiteboard sketches.

    \section{Visual Coding Assistant \label{sec:inkai}}

To study these phenomena, we developed a VSCode plugin where users can open a snapshot of a whiteboard sketch and have a Jupyter Notebook generated from it. This tool acts as part of the experimental materials and is used to test part of our theory regarding the usefulness of Visual Coding Assistants. In this section, we provide details about the prototype implementation and a motivating example.

\textbf{Interaction.} We want participants to focus on sketching and the conceptualization of in-IDE Visual Coding Assistants. The interaction that we seek to study is the naturalness of information exchange between the developer and the AI using the drawing in the same way a developer would do with other humans. We are not interested in studying the plug-in user interface interactions, leaving this factor open to future research.
Therefore, our VSCode extension simplifies the user interaction to a single button that generates code. When a picture is opened in the IDE, the plugin detects the image and allows the user to click on the \menu{Generate Code} button. Otherwise, a warning will tell the user that the open file is not an image, and therefore cannot be used to generate code.
Once the user clicks on the button, a new panel is opened to the right of the active window, which displays the generated Jupyter Notebook.

\textbf{Code Generation} The code generation is delegated to OpenAI's GPT-4o with vision capabilities by using their web API. At the time the study was conducted, the model performs better than any other regarding Vision capabilities according to the benchmarks updated daily at HuggingFace \cite{yujie2024wildvisionarena, 2023opencompass}.\footnote{\url{https://huggingface.co/spaces/WildVision/vision-arena} \& \\ \url{https://huggingface.co/spaces/opencompass/open_vlm_leaderboard} \\ Accessed: May 27th 2024} GPT4 is preferred both for the convenience of using the API and its performance.
The prompt used in the final version of the prototype and the user study is the following:

\begin{promptbox}    
\textit{You are an expert code generation system and you are specialist in Python.
You will receive an image representing a ML workflow and you will reply with the correspondent steps in JSON list format.
The output for each step must have exactly two elements: ``code'' field with Python code;
and ``markdown'' field containing the explanation. Markdown explanations must be short and easy to understand.
No introductory text is allowed in the output.
Retrieve only the list format, e.g.[\{code: ``...'', markdown:``...''\}, \{code: ``...'', markdown: ``...'' \}] \\
Create the code for the following diagram.
}
\end{promptbox}

\textbf{Assembling the Notebook} After receiving the code generated from the model, the plugin gathers all the identified steps and assembles them into a Jupyter Notebook. Each step in the image is then translated into a pair of cells - a markdown cell that provides a brief explanation, followed by the code cell that contains the code to be executed.

\subsection{Motivating Example}
\begin{figure*}[ht]

\centering
\includegraphics[width=\textwidth]{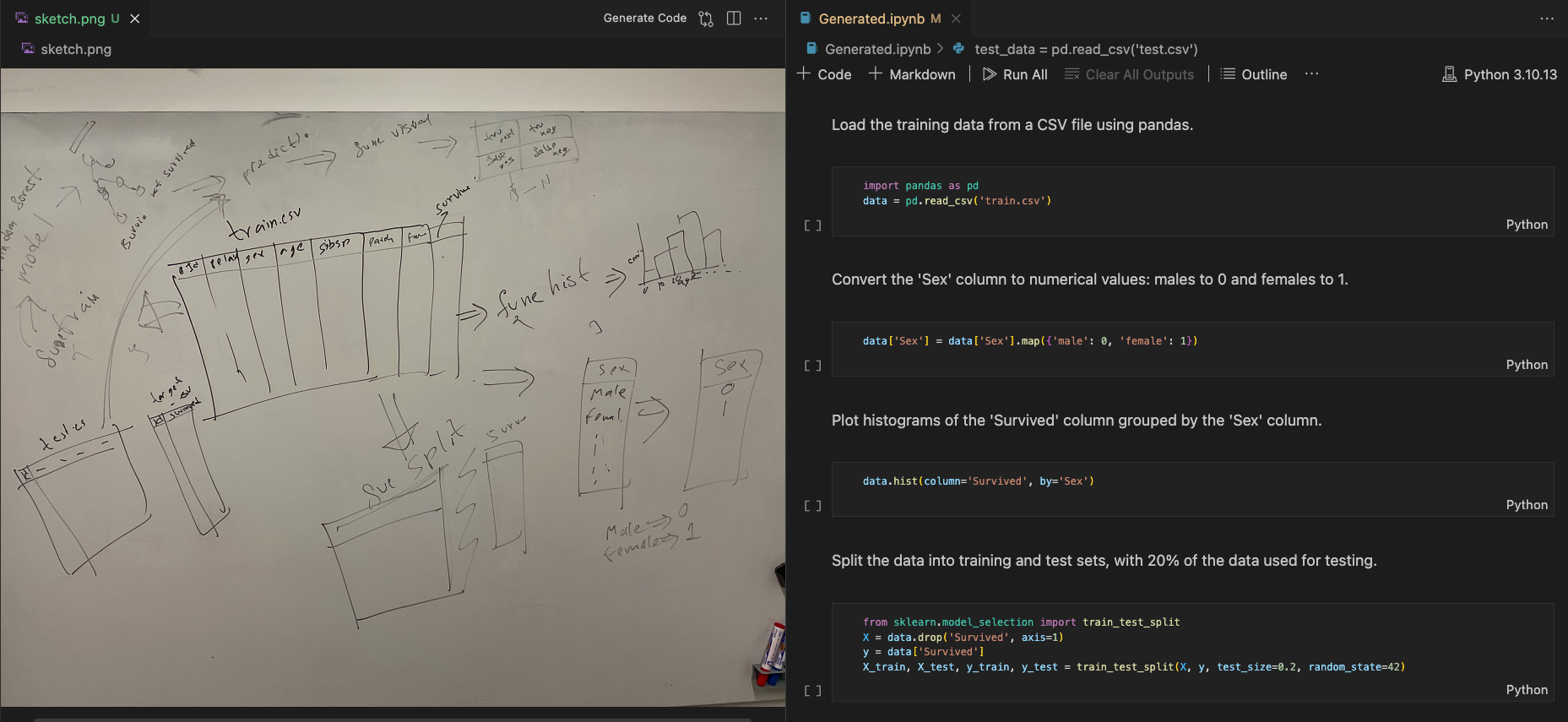}
\caption{Example of usage of our prototype. Left editor tab: The whiteboard sketch uploaded by the participant to be used by the code assistant to generate code. Right editor tab: Generated Jupyter Notebook with code and markdown explanations.}
\label{fig:example}
\end{figure*} 

This study recreates situations where developers want to communicate their ideas using the whiteboard. We give a task to each participant and ask them to behave as in a regular meeting where they explain how to solve it. After sketching, they use our prototype to generate code. Figure \ref{fig:example} shows an example of one of our participants' usage of the prototype during the study. 
In this example, the participant explains how they want to load three csv files (``test.csv'', ``target.csv'' and ``test.csv''), visualize ``age" with a histogram, map ``Sex'' (Male/Female) to numerical variables (0/1), train a Random Forest model and use a confusion matrix to see the results. The participant uses a diagram to represent the precedence of each operation. In the center, we have a data table that needs to be loaded from the ``train.csv'' file. Most of the other operations depend on this one, so it must occur before the others. Causal dependencies are indicated with arrows. The participant represents the singular operations using a mix of visual icons, such as the representation of the histogram, and many short, textual annotations, such as ``$Male \Rightarrow 0$''.

The user uploads a picture of their whiteboard sketch to the repository they are working on and after clicking the ``Generate Code'' button, a Jupyter Notebook is created and opened next to the image.
Using the drawing as the only source for code generation, we see the Code Assistant preserving the order of operations, while maintaining a coherent succession of steps. At the same time, it understands most of the participant intentions, such as the correct mapping, the use of a histogram function or the split of the variable ``Survived'' from the dataset. 
On the other hand, some inaccuracies are visible, such as the arguments to the histogram function or splitting the train dataset into train and test. These are things the developer may want to change or remove, according to their initial intention.
In real scenarios, even with the inherent inaccuracies, this is useful to quickly generate a first code skeleton and discuss ideas at the code level during the meeting. It saves time by reducing the time on programming activities and making the design clearer to everyone in the meeting.

\section{Methodology}
The goal of our research study is to investigate developers' behavior while using the whiteboard and Visual Code Assistants. Our population of interest is individuals whose needs encompass ML and DS tasks, independent of their level of knowledge or expertise in this domain. Our sample comprises individuals from academia and industry, who have previously worked with Python and understand ML and DS concepts.

\subsection{Recruitment} We recruited participants in two ways. The first was by identifying researchers and developers from personal contacts and contacting them directly via email. The candidates were selected taking into account previous experience. The second was by employing snowball sampling, asking previous participants to nominate new people who meet the criteria and they believed would accept to take part in the study. Each participant was compensated with a \$20 Amazon gift card. We recruited 19 participants (11 female, 7 male, 1 non-binary) from different institutions in industry (4 participants - Meta, Ansys, TripAdvisor, Bandora) and academia (15 participants - CMU, NYU, Univ. of Porto, Univ. of Lisbon, Univ. of Waterloo, Brown Univ., TU Delft, Bowdoin College, Colby College, Bucknell University). Regarding the educational levels in our sample, 10 participants were pursuing or had completed a doctorate, five a master's degree, and four a bachelor's degree. Regarding their current occupation, four are working in ML or DS as their main activity, nine in Software Engineering or other programming-intensive fields, and six in other fields.

\subsection{Protocol}
Our IRB-approved protocol encompasses five stages after the participant agrees to take part in the study: 1) Pre-Task Survey, 2) Sketching, 3) Coding, 4) Interview and 5) Post-Task Survey.
We designed the study to take a maximum of 60 minutes per participant. Stages 1 and 5 occur before and after the study, respectively. The study was conducted in person or virtually, depending on the participant's availability. 

A pilot study with five participants was conducted to improve and validate the protocol and the tasks used in the final study. In this section, we detail each of the parts. 

\textbf{Pre-Task Survey.} Participants complete a survey collecting demographic information and details about their sketching and coding habits. This information is crucial for understanding the diverse backgrounds and experiences of the participants, which can influence their interactions and measurements during the study. The survey is hosted at Google Forms and a link is sent to the participant before the study.

\textbf{Sketching.} Each participant is presented with an ML task and given a few minutes to get familiarized with the task and dataset. We allow the participant to ask clarifying questions about the task before starting the sketch. When participants confirm that they understand the task, they are asked to impersonate the team member responsible for solving the problem and simulate a whiteboard meeting where they explain their intended solution to an ML developer. At this point, a timer is started to capture sketching time, and audio is recorded to capture their thought process (think aloud). 
Participants were instructed to draw until they felt their explanation was complete. At this point, the timer and recording were stopped, and we captured a picture of the final sketch.

\textbf{Coding}: A break is made before proceeding to the coding stage. The break allows participants to switch to a new context since now they impersonate an ML developer and simulate a programming session. In the meantime we set up the in-browser VS Code IDE using GitHub CodeSpaces. This workspace includes the Visual Coding Assistant prototype presented in section \ref{sec:inkai} and common ML libraries that can be useful to solve the tasks. The drawing is uploaded to this workspace, allowing the user to generate code from it. Their sketch is used to generate code using the prototype and a Jupyter Notebook with the same base name as the image is generated. The participant renames the file to ``generated.ipynb'', and creates a copy of it calling it ``solution.ipynb''. The participant can make changes to the ``solution'' file until they are ready to submit their final solution. During the code session, a timer is started, and audio and screen are recorded through Zoom for further analysis. The think-aloud protocol is used and the participant is encouraged to speak while they do the task. During the coding session, the participant can utilize all tools that they would use in a real situation, such as Google or ChatGPT. Interactions with such tools are registered as well. To ensure the user study remains within the 60-minute time slot, participants are assigned 30 minutes to complete this step. To capture real scenario patterns, participants are not informed of this time limit beforehand, preventing any bias towards utilizing the entire 30 minutes.

\textbf{Interview Questions.} 
We conclude the user study with a semi-structured interview, focusing on three key areas to understand developers' needs regarding Visual Coding Assistants. The first area addresses sketching, discussing the intuitiveness of expressing coding ideas as sketches and how sketching patterns may influence code generation. The second area involves the participant reconstructing their work environment to identify potential use cases where they could benefit from this technology. The final area explores future directions for developing Coding Assistants with vision capabilities and gathering participant feedback on potential features and improvements.

\textbf{Post-Task Survey.} Participants are asked to fill out a survey designed to collect feedback on various aspects of the Visual Coding Assistant tool. The Google Forms link is sent to the participant after the user study.
The survey begins by assessing participants' initial confidence in the tool's ability to generate accurate code from their sketches. Participants rate the performance of the tool, providing their overall satisfaction with the accuracy of the generated code, and the frequency of discrepancies between their intentions and the code that was generated.
Additionally, participants evaluate the tool's usefulness and provide their perceptions of how code generation from sketching could impact their productivity.

\subsection{Tasks}  
We designed the tasks to simulate realistic DS scenarios in which developers want to express their ideas visually before coding, e.g., during whiteboard meetings. 
The tasks are similar to those encountered in ML course assignments and coding challenges. To increase the generalizability of the results, we designed three ML tasks that center around typical modeling settings: binary classification, regression, and image classification. Each task has the same types and number of subtasks, such as data loading and visualization. Table \ref{tab:tasks_subtasks} shows a summary of the three tasks that are used in the user study.

\begin{table}[h!]
\centering
\caption{Tasks and Subtasks}
\label{tab:tasks_subtasks}
\begin{tabular}{|l|l|l|l|}
\hline
\multicolumn{1}{|c|}{\textbf{SubTask}} & \multicolumn{1}{c|}{Task1}                                                & \multicolumn{1}{c|}{Task2}                                                 & \multicolumn{1}{c|}{Task3}                                     \\ \hline
\textbf{Load}                          & \begin{tabular}[c]{@{}l@{}}Load one \\ CSV file.\end{tabular}             & \begin{tabular}[c]{@{}l@{}}Load three\\  CSV files.\end{tabular}           & \begin{tabular}[c]{@{}l@{}}Load two \\ CSV files.\end{tabular} \\ \hline
\textbf{Transform}                       & \begin{tabular}[c]{@{}l@{}}Split target. \\ Split train/test\end{tabular} & \begin{tabular}[c]{@{}l@{}}Split target. \\ Variable encoding\end{tabular} & Reshape image                                                  \\ \hline
\textbf{Plot1}                       & Boxplot                                                                   & Histogram                                                                  & Grid Plot                                                      \\ \hline
\textbf{Plot2}                       & Scatter Plot                                                              & Confusion Matrix                                                           & Line Plot                                                      \\ \hline
\textbf{Model}                         & Linear Regression                                                         & Random Forest                                                              & CNN                                                            \\ \hline
\end{tabular}
\end{table}
Each task was designed by starting from a Kaggle dataset and defining the five subtasks afterward. The subtasks were chosen taking into account common DS and ML routines. 

The datasets were modified to guarantee that no extra complexity was added to the task and no extra execution time was needed, making it feasible to complete it in the stipulated time frame (1 hour).
For Task 1, we use the House Prices dataset (\url{https://www.kaggle.com/datasets/vikrishnan/boston-house-prices}), for Task 2 the Titanic dataset (\url{https://www.kaggle.com/c/titanic/data}) and for Task 3 the Digit Recognition dataset (\url{https://www.kaggle.com/competitions/digit-recognizer/data}).

Tasks were assigned randomly for each participant, following a round-robin approach, with assignments made upon participant acceptance into the study. This procedure ensures a balanced distribution of tasks, with an identical number of participants per task.

\subsection{Data and Artifacts}
During the user study, we collect both qualitative and quantitative data. For privacy reasons, all the recordings and transcripts are not publicly available. This includes the screen recording of the coding sessions, and the audio recordings with the corresponding transcripts of the sketch sessions, coding sessions, and interviews. 

In the publicly available data, alongside the original sketches and the generated Jupyter Notebooks, we include coding session annotations, interview thematic analysis, and sketching thematic analysis documents. The raw data of pre-task and post-task surveys is also available, after anonymizing participants' identities. All data and scripts can be found at \url{https://zenodo.org/doi/10.5281/zenodo.13184028}.

\subsection{Evaluation}
Our evaluation involves both qualitative and quantitative methods. In this exploratory study, we follow a constructivist paradigm \cite{easterbrook2008selecting}, where we collect data and draw conclusions from it in an inductive manner. This section details the methodology used to answer each RQ in this study.

\subsubsection{Sketch Analysis}
In our study, we employed conceptual modeling \cite{singer2008software} alongside the think-aloud protocol \cite{Ericsson1993} to gather DS sketches. Participants were asked to simulate a whiteboard meeting, where they explained their approach to a given task using a whiteboard, paper, or a tablet. To analyze these sketches, we first applied open coding to identify and categorize recurring characteristics. This was followed by axial coding, which allowed us to consolidate the findings into two main categories, detailed in the results section.

Due to some inherent subjectivity in interpreting the sketches, our analysis was further supported by the audio recordings and transcripts from the sketching sessions. This additional data enabled us to trace the connection between specific subtasks and corresponding parts of the drawings, providing a clearer understanding of the sketching process and enhancing the reliability of our findings.

\subsubsection{Code Generation}
Sketching is inherently subjective, which makes the development of metrics for evaluating code generated from another person's drawings harder. Given the inherent limitations of visual representations, which often omit crucial details, it is unrealistic to expect the LLM to perfectly interpret and translate all aspects of a sketch into accurate code. To address this challenge, we established a controlled environment where participants follow specific tasks to guide their sketches. This setup enables us to create metrics based on these tasks and, by extension, infer the accuracy of the generated and submitted code. We rely on one \textbf{key assumption:} \textit{Participants correctly comprehend and execute the provided tasks in their sketches.} This assumption is supported by audio recordings and transcripts of the sketching sessions, which are used during the sketch analysis.

\begin{figure*}
\centering
\includegraphics[width=\textwidth]{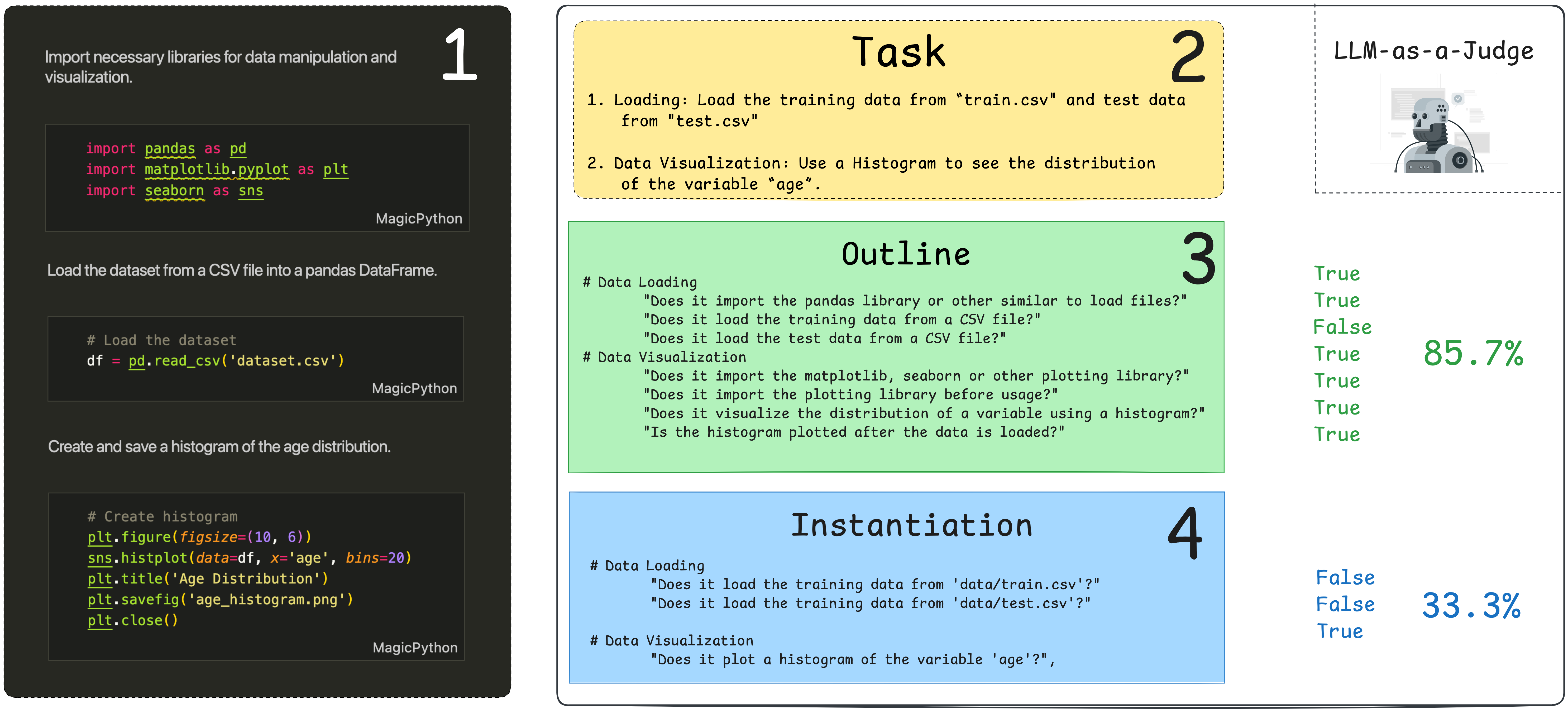}
\caption{Overview of our benchmarking using LLM-as-a-Judge. To evaluate a Jupyter Notebook generated from a sketch (1), the Judge has access to knowledge elements related to the original drawing: a Task description that relates to users' coding intentions depicted on the drawing (2), and the grading criteria for the Outline and Instantiation metrics (3 and 4).}
\label{fig:llm_judge}
\end{figure*}

\subsubsection{LLM-as-a-Judge}
To calculate the metrics, we employ a code analysis strategy termed LLM-as-a-Judge. This approach leverages a large language model (LLM) to evaluate code based on predefined criteria, mirroring how a human evaluator would grade notebooks. The primary motivation for using LLM-as-a-Judge rather than manual grading is to ensure consistent and efficient evaluation across the dataset, particularly as new notebooks are added. Manual grading is time-intensive and lacks scalability, becoming increasingly impractical for larger datasets. In prior work, LLM-as-judge approaches achieved agreement levels comparable to human evaluators in previous work \cite{zheng2023judgingllm} and are widely used in settings like Chatbot Arena \cite{chiang2024chatbot}. We used GPT-4 as the judge for pointwise grading as it has been reported to perform well in such task \cite{huang2024limitationsfinetunedjudgemodels}. A simplified example of the evaluation process is illustrated in Figure \ref{fig:llm_judge}. An evaluation of this judge in our setting is provided below.

Although this is not as precise as formal program analysis, it is more flexible in task presentation and evaluation. Participants in our study may use a wide range of libraries, e.g., pandas, csv, numpy, or dask for file loading. Implementing formal analyzers to check the correctness of pipelines would be prohibitively expensive without strict restrictions on the libraries and functions participants can use, which would reduce the realism of our study. Execution-level metrics are another common choice \cite{chen2021evaluating}, but require fully functional notebooks. Notebooks generated from sketches are virtually never fully functional, nor is it feasible to specify complete test cases for them. Given this, we opted for the flexibility of using an LLM as a judge, combined with detailed specifications.

During the development of the LLM judge, ground-truth Jupyter Notebooks for each task were used to verify how well the model performed against a perfect solution. For each subtask, specific parts of the code were manually removed to ensure the LLM correctly flagged the corresponding questions as `False'. This process allowed us to curate a better evaluator, improving both the prompt and the ``grading'' questions. The final version of the prompt used in the evaluator was the following:

\begin{promptbox}
    You are a programming teacher responsible for grading Data Science coding assignments. \\
You receive a Jupyter Notebook to be graded and a set of questions that you should answer with True or False. \\
You should return the same list of questions answered with the corresponding answer `True' or `False'. \\
For example, the output should be: Question1: True \textbackslash n Question2: False\textbackslash n etc.
\end{promptbox}

\textbf{Judge Validation.} During the study, a second evaluation was performed, manually grading and comparing the results with the judge of a subsample of the notebooks submitted by study participants. To avoid confusion with the results presented in section \ref{sec:results}, we refer to \textit{judgment} metrics as the ones that are used to evaluate how well the LLM-as-a-Judge is capable of assessing the correctness of a Jupyter Notebook. These are intrinsic to the specific LLM that is used as the judge.

The Outline evaluation demonstrated a judgment accuracy of 92.6\% (63/68), and a judgment precision of 92.2\% (59/64). The Instantiation evaluation showed a judgment accuracy of 78.3\% (36/46) and a judgment precision of 74.3\% (26/35). These results suggest that the LLM-as-Judge methodology is reliable for this use case, as it demonstrates high levels of agreement with human judgment.

The \textit{generation} metrics are specific to each input notebook, as shown in the example of Figure \ref{fig:llm_judge}. These are reported in section \ref{sec:results}.

Our quantitative benchmark evaluates Jupyter Notebooks based on the following metrics. \textbf{Outline Accuracy}: the percentage of correctly identified high-level steps in the notebook according to the sketch outline. It focuses on whether the notebook includes logic for all major sub-tasks, regardless of their specific implementation details. \textbf{Instantiation Accuracy:} the percentage of correctly implemented details within the notebook based on the sketch. This captures whether the specific elements or parameters match those sketched by the user. The instantiation accuracy is a sub-specification of the outline metric, and they are correlated. Models that perform well in the instantiation metric will perform well in the outline, but the reverse may not be true. The reason for having two different metrics is to understand the degree of detail captured by the vision capabilities of LLMs. This interacts with the sketches themselves: they virtually always contain the details required to construct the outline, but may omit details required to implement the code completely, such as the input file name. Hence, we typically do not expect a perfect instantiation accuracy.

While our main study uses GPT-4o, we conduct a retrospective analysis of three models on the collected sketches: Claude 3.5 Sonnet, Gemini Pro 1.5, and OpenAI’s GPT-4o. For each model, we feed the same set of sketches, generating a Jupyter Notebook per sketch. We compare both the outline and instantiation performance. 
Please note that GPT-4o serves as both the judge and a participant in this experiment. While this might introduce bias, the judge only has access to the generated notebook, the task, and the grading criteria, not the model name, so we expect that the risk of bias is slim.

\subsubsection{Surveys and Regression Analysis}

We use our benchmark to evaluate the code generated from sketches, analyzing the relationship between sketching time and code accuracy through regression analysis. This analysis considers participants' programming experience and sketching expertise, which we collected from our pre-task survey. We also control for various levels of programming knowledge and sketching expertise to understand their impact on code accuracy. Other variables from the pre-task survey were used primarily for demographic reporting, as detailed in the recruitment section. Similarly, data from the post-task survey provides additional evidence that supports our qualitative results and complements the insights gathered from the interviews.

\subsubsection{Interview Analysis}

The approach to analyzing interview data closely follows the methodology employed in sketch analysis. Initially, open coding is applied to the transcripts to identify and categorize the raw data. This preliminary step involves breaking down the text into discrete units and labeling them with codes that represent significant observations.
Subsequent to open coding, axial coding is utilized to discern and establish relationships between these codes. This process helps to organize the data into coherent categories that address specific themes such as emotional responses, sketching practices, coding behaviors, and future perspectives. 

Building upon these categories, an inductive thematic analysis is conducted. By identifying recurring patterns and thematic elements, this approach helps to summarize the findings into a comprehensive report that reflects the participants' viewpoints.
To measure the robustness of the findings, the percentage of interviewees mentioning each theme is calculated. This quantitative measure provides insight into the preponderance of certain perspectives among participants, highlighting the strength of our conclusions. 
The audio and video recordings of the sketch and coding session are used as support during the analysis of the sketches and as confirmation of the coding session annotations, such as the number of times participants assessed ChatGPT.

\section{Results \label{sec:results}}
\subsection{How Do Data Scientists Sketch?}

To effectively design and implement visual methods in Code Assistants, understanding prevalent whiteboard sketching practices is essential. This section explores the findings that address \textbf{RQ1: What are the prevalent sketching patterns of ML developers?}

\textbf{Sketching Practices: Quantitative Findings.} In our pre-task survey, we assessed the last time participants used sketching to express their mental models during work activities. The results indicate that 36.8\% of participants used sketching within the last week, 26.3\% within the last month, and 21.1\% within the last three months. Additionally, 15.8\% had either not used sketching in over a year or did not recall their last use. These results suggest that sketching is actively used by a substantial portion of participants, with 63.1\% engaging in it within the last month alone.

\textbf{Sketching Practices: Qualitative Insights.} This finding aligns with interview insights that illustrate varied attitudes toward sketching. Participants who do not sketch often expressed a preference for having control over their code directly or using text prompts. P1 mentioned, ``\textit{[Sketching] is something that you don't do very frequently, you don't need to do it when you have a data set that you want to look at}''. 
P13 and P17 show their preference for text prompts, ``\textit{I would be tempted to use the text [prompts] more often.''} and ``\textit{I don't trust a computer [...] to recognize sketches. So, I prefer ChatGPT.}'', respectively.
Conversely, participants who sketch frequently cited its utility in clarifying and structuring their thoughts. P7 stated, ``\textit{Sometimes, I do the drawings for my benefit so that I don't get as confused}'' and P12 elaborated, ``\textit{When you write down you have a whole structure [...] when you are doing the sketches, it's like you already run the whole thing in your mind.}''. 
This suggests that sketching serves as a cognitive aid, helping developers to visualize and organize their ideas before coding, as an integral part of their problem-solving process. Along the same lines, some participants expressed their preference for the new possibility presented during the study of using sketch prompts to generate code. P9 observed that ``\textit{[Sketching] is faster than writing the code for sure},'' and P11 found sketching to be more efficient compared to detailed prompt explanations, noting, ``\textit{[Prompting] also requires you more time to explain. So just being able to sketch some stuff [...] and generate the code was interesting to me}.'' This preference aligns with the survey data showing that a significant portion of participants use sketching regularly, suggesting it is valued for its speed and clarity in coding tasks.

\begin{mainbox}{Finding 1 - Sketching as a Development Tool} 
Developers commonly leverage sketching as a cognitive tool to visualize and organize ideas during the problem-solving process, with 63.1\% of developers engaging in sketching within the past month. While some prefer traditional text-based prompts or direct coding approaches, \textbf{there is a significant demand for tools that can directly translate sketches into code}.
\end{mainbox}

\textbf{Sketch Styles.} The first finding tells us \textit{why} developers sketch. To understand \textit{how} they sketch, we analyzed the sketches acquired during the study. As our focus is on people working in DS, the patterns we observed are linked to ML and data-related tasks. 
While it is expected given the nature of the tasks, a notable finding from our analysis is the strong emphasis developers place on maintaining order in their drawings. Participants exhibited clear mental models for structuring task dependencies, which were primarily expressed through visual representations. The following structural representations were found in the sketches:\footnote{Note that multiple constructs can be combined on the same sketch so the percentages add up to more than 100\%.}

\textbf{Diagrams}: The most common approach (52.6\%) was to use diagrams to represent relationships. Participants typically employed arrows to indicate dependencies, data flow, or the sequence of operations. These diagrams often exhibited a spatial orientation, such as left-to-right or top-to-bottom progression. Nodes within these diagrams represented components, objects, entities, or groups of operations.
\\ \noindent \textbf{Sequential Lists}: While less frequent than diagrams (42.1\%), some participants used sequential lists to outline task order. These lists relied on spatial arrangement (e.g., top to bottom) to imply ordering.
\\ \noindent \textbf{Numbering}: In some cases (36.8\%), participants used numbers to clarify the order of operations
\\ \noindent \textbf{Freeform}: Occasionally (10\%), sketches lacked explicit structural elements, suggesting that participants were focusing on capturing ideas without immediate concern for order.

From the sketch analysis, we further identified patterns regarding how participants represented the operations within each subtask. Developers combine the following five elements:
\\ \noindent \textbf{Boxes:} Participants frequently employed boxes or other enclosed shapes to delineate distinct operational units within their sketches. These boxes are often combined with labels or icons.
\\ \noindent  \textbf{Iconography:} Symbolic or graphical representations of data or concepts.
\\ \noindent  \textbf{Annotations:} Short text elements providing essential context for understanding the intended operations. 
\\ \noindent  \textbf{Descriptions:} Longer and more descriptive textual explanations of operations, often using verbs.
\\ \noindent  \textbf{Code:} Python code, pseudo-code, or code-like expressions.

\begin{table}[]

\caption{\label{tab:elem_subtask} Usage of element by subtask. Textual annotations are the most commonly used element in all tasks.} 
\centering
\begin{tabular}{l|r|r|r|r|r|}
\cline{2-6}
                                           & \multicolumn{1}{l|}{\textbf{Load}} & \multicolumn{1}{l|}{\textbf{Plot1}} & \multicolumn{1}{l|}{\textbf{Transform}} & \multicolumn{1}{l|}{\textbf{Model}} & \multicolumn{1}{l|}{\textbf{Plot2}} \\ \hline
\multicolumn{1}{|l|}{\textbf{Box}}         & 7                                  & 8                                   & 9                                       & 8                                   & 7                                   \\ \hline
\multicolumn{1}{|l|}{\textbf{Iconography}} & 6                                  & 10                                  & 6                                       & 4                                   & 9                                   \\ \hline
\multicolumn{1}{|l|}{\textbf{Annotation}}  & \textbf{13}                        & \textbf{14}                         & \textbf{15}                             & \textbf{11}                         & \textbf{14}                         \\ \hline
\multicolumn{1}{|l|}{\textbf{Description}} & 2                                  & 1                                   & 2                                       & 4                                   & 3                                   \\ \hline
\multicolumn{1}{|l|}{\textbf{Code}}        & 6                                  & 6                                   & 4                                       & 6                                   & 3                                   \\ \hline
\end{tabular}
\end{table}

As illustrated in Table \ref{tab:elem_subtask}, textual annotations were the most frequently employed element across all subtask types, indicating a strong preference for textual descriptions of operations. While both Plot1 and Plot2 subtasks exhibited a pronounced reliance on iconography, suggesting a potential advantage of visual representations for data visualization, the creation of machine learning models was primarily conveyed through detailed textual descriptions. This was also mentioned during interviews. For instance, P4 mentioned ``\textit{[Sketching] is more intuitive than having to look up function calls on the APIs [...] there are different variations [of plots] and it's not exactly what I wanted visually.}'' This also highlights that image- and text-based code generation are not mutually exclusive.

\begin{mainbox}{Finding 2 - Structure, Text and Icons} 
    Developers use the whiteboard to organize their ideas by maintaining an order of operations.
    They rely on \textbf{diverse types of elements, including textual annotations for details and icons for visual tasks.} Effective visual assistants must prioritize accurate text recognition, while also considering spatial layout and visual symbols for enhanced comprehension.
\end{mainbox}

\subsection{Sketch to Code Generation}

In this section, we examine the results regarding \textbf{RQ2: To what extent can current sketch-to-code technologies support ML developers?} Our focus is on understanding the complexities of transforming whiteboard sketches into executable code, exploring developers' expectations for such tools, and assessing the capabilities of current large language models in this domain.

\textbf{Results.} On average, participants spent 9 minutes sketching (min: 2, max: 15) and 16 minutes programming (min: 4, max: 37\footnote{An outlier; we instated the 30 minute coding time cap mentioned in the methodology after one participant spent 37 minutes on the coding portion.}). 
This average sketch yields code with an outline accuracy of 79\% and an instantiation accuracy of 36\%. The median number of lines changed starting from the generated code in the final solution is 19.5, where the median solution has 40 lines in total. Given that, \textbf{sketching reduced the amount of written lines of code by 49\%} on average. 

To support this analysis, a manual verification of the results obtained from the submitted notebooks was conducted, with each subtask evaluated as either correct or incorrect. Subtasks were considered correct if they were executed successfully and produced the expected output. In total, 11 out of 19 participants successfully solved their assigned challenge, while an additional 3 participants completed 4 out of 5 subtasks correctly.

\textbf{Individual Differences.} To investigate the relationship between sketch characteristics and code generation outcomes, we conducted linear regression analyses to predict outline and instantiation accuracies. Three key variables were considered.
First, we examined \textbf{sketch time}, the duration of the whiteboard sketching process, measured in minutes. Second, we explored \textbf{last sketching}, a proxy for sketching proficiency, representing the time elapsed since the participant's last sketching activity. This variable was included under the assumption that participants who had recently engaged in sketching activities would likely exhibit better drawing skills. Finally, we controlled for \textbf{programming experience}, measured in years.

\begin{table}[t]
\caption{Summary of regression models predicting code generation outcomes based on sketch time, last sketching activity, and programming experience. Reported p-Values are after Benjamini-Hochberg correction.}
\label{tab:linreg_sketch}
\centering
\begin{tabular}{c|cc|cc|}
\cline{2-5}
                                                          & \multicolumn{2}{c|}{\textbf{Instantiation}}             & \multicolumn{2}{c|}{\textbf{Outline}}                   \\ \hline
\multicolumn{1}{|c|}{\textbf{Predictor}}                  & \multicolumn{1}{c|}{\textbf{Coeff.}} & \textbf{p-Value} & \multicolumn{1}{c|}{\textbf{Coeff.}} & \textbf{p-Value} \\ \hline
\multicolumn{1}{|c|}{Intercept}                           & \multicolumn{1}{c|}{24.5364}         & 0.105            & \multicolumn{1}{c|}{55.9498}         & 0.004**            \\ \hline
\multicolumn{1}{|c|}{Sketch Time}                         & \multicolumn{1}{c|}{1.4536}          & 0.284            & \multicolumn{1}{c|}{2.4678}          & 0.105            \\ \hline
\multicolumn{1}{|c|}{Last Sketching}                      & \multicolumn{1}{c|}{0.0266}          & 0.686            & \multicolumn{1}{c|}{0.0112}          & 0.883            \\ \hline
\multicolumn{1}{|c|}{Experience}                          & \multicolumn{1}{c|}{-0.4327}         & 0.826            & \multicolumn{1}{c|}{0.0310}          & 0.976            \\ \hline
\multicolumn{1}{|c|}{$R^2$}                               & \multicolumn{2}{c|}{0.156}                              & \multicolumn{2}{c|}{0.297}                              \\ \hline
\multicolumn{1}{|c|}{\textit{Adj} $R^2$} & \multicolumn{2}{c|}{-0.025}                             & \multicolumn{2}{c|}{0.146}                              \\ \hline
\end{tabular}
\begin{flushright}
\textit{P-Values:} (*) $p < 0.1$; (**) $p < 0.05$; (***) $p < 0.01$ \\
\textbf{Adjusted p-values for multiple testing (B-H correction)}
\end{flushright}

\end{table}

Table \ref{tab:linreg_sketch} presents the results of the linear regression analyses predicting code instantiation and outline generation. The p-values are adjusted using the Benjamini-Hochberg correction to control the false discovery rate and account for multiple hypothesis testing. While neither sketch time, last sketching activity, nor programming experience significantly influenced code instantiation accuracy, the model for outline generation revealed a positive relationship between sketch time and generation accuracy. Specifically, each additional minute spent sketching was associated with a 2.4 percentage point increase in outline accuracy, controlling for other variables. The p-value for this variable (0.105) lies slightly outside the standard confidence intervals (0.1), which indicates that a larger sample size is needed to better estimate the true coefficient of this variable.
This corresponds to an average outline accuracy of ca. 75\% for sketches completed in 7 minutes to ca. 85\% for ones completed in 11 minutes.

Worth noting that the intercept in the outline generation model indicates a baseline outline accuracy of 55.9\% when the other predictors are zero. This does not imply that a completely blank sketch would generate any code. Rather, it suggests that minimal sketching efforts could lead to a reasonably accurate code outline. For instance, one participant's sketch made in under four minutes reached an outline accuracy of 90\%.

\begin{mainbox}{Finding 3 - Sketch Duration and Accuracy} 
    \textbf{Even brief sketching periods have the potential to generate useful code outlines.} Longer sketch durations are associated with improved outline accuracy. This suggests that code generation tools can be effectively adapted to user needs based on the desired level of code generation detail.
\end{mainbox}

\textbf{Model Comparison.} In our prototype we used GPT-4o as the underlying model. We evaluate the performance of other leading commercial alternatives with similar functionalities, including vision capabilities and public APIs. The alternatives that we compare are Google’s Gemini 1.5 Pro, Anthropic’s Claude 3.5 Sonnet, and OpenAI’s GPT-4o.

\begin{figure}[t]
    \centering
    \includegraphics[width=.9\linewidth] {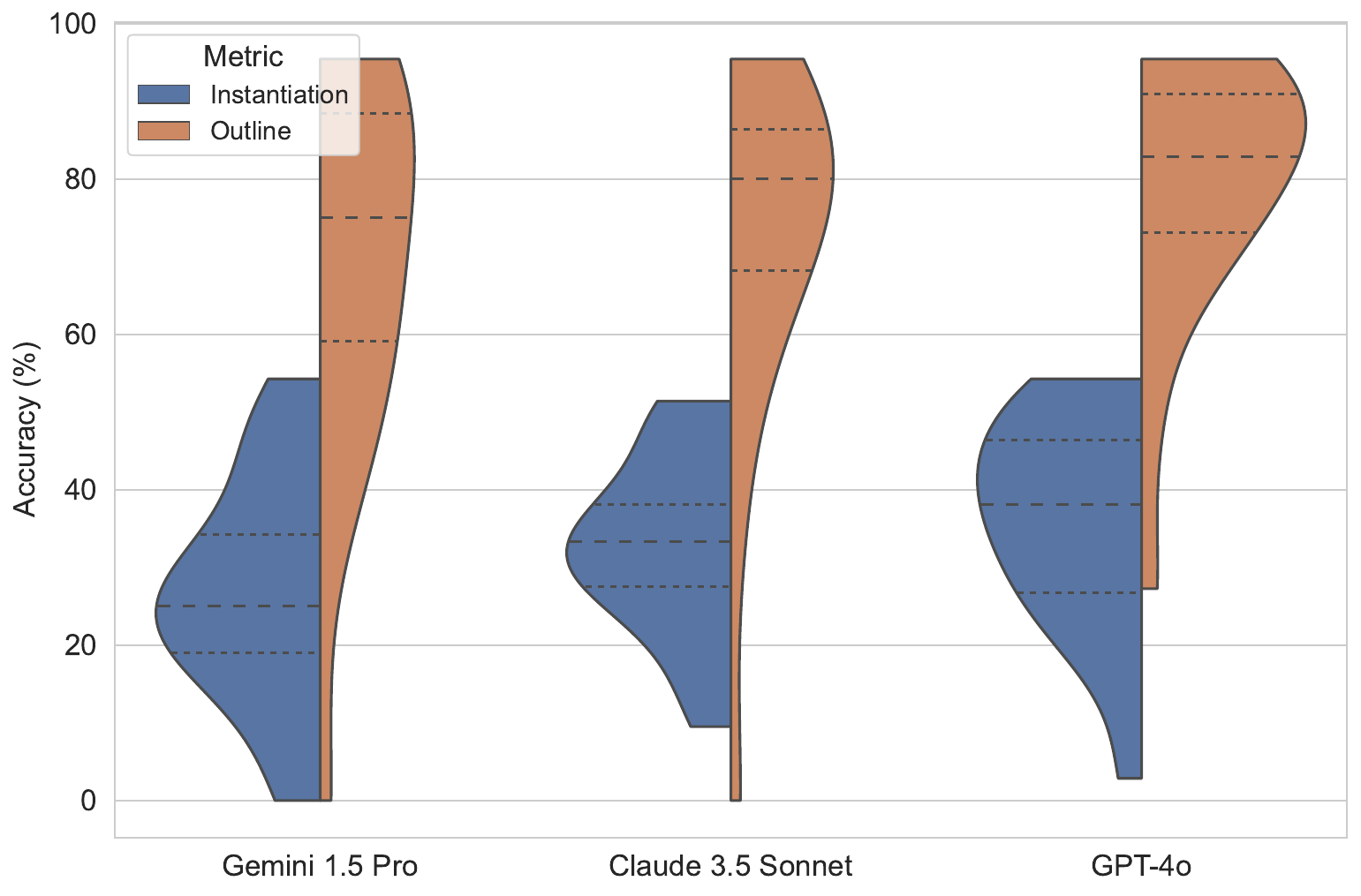} 
\caption{Performance comparison of top visual LLMs on the 19 notebooks. No statistically significant difference was detected in either dimension in our study.}
\label{fig:models_performance}
\vspace{-.2cm}
\end{figure}

Figure \ref{fig:models_performance} shows the results of the three models on our two metrics automatically calculated using the LLM-as-a-Judge approach. Outline accuracy for GPT-4o, Claude, and Gemini is 79\%, 71\%, and 68\%, respectively. Similarly, GPT-4o leads in instantiation accuracy with a score of 36\%, followed by Claude at 33\% and Gemini at 26\%.
While statistical significance was not found in code generation performance among the models, there may still be some evidence of differences worth further exploration with a larger dataset.

\begin{mainbox}{Finding 4 - Generation Across Models}
    \textbf{There is no significant difference in accuracy between GPT-4o, Gemini 1.5 Pro, and Claude 3.5 Sonnet.}
All three models performed substantially better at generating code skeletons (outline) compared to handling detailed code implementations (instantiation).

\end{mainbox}

\vspace{-.2cm}
\subsection{Perspectives on Visual Code Assistants}

To maximize the utility of these models, it's crucial to integrate them into tools that assist developers in real-world scenarios.   The concept of a Visual Code Assistant involves embedding vision capabilities into existing code assistants like GitHub Copilot. To understand the potential of this approach, we investigate developers' needs and expectations through
\textbf{RQ3: What are developers' perspectives on the value and potential of in-IDE Visual Code Assistants?}

Participants expressed a desire for a few key characteristics in visual code assistants. Explainability and guidelines were frequently mentioned, with users seeking to understand the rationale behind generated code and how to effectively sketch and utilize the tool. Some participants show their contentment about the generated markdown explanations: ``\textit{[it] was the most impressing part because already give these parts here that are not code, but just information}'' (P3).
On the other side, some participants wished they had more information, like 
P7 who captured the sentiment: ``\textit{if there are parts that don't require as much detail, I think those are important guidelines for the users. Like, this much detail helps the model generate better code}''.

Iterative and interactive code generation was identified as another important aspect. Participants emphasized the dynamic nature of the programming process, highlighting the need to modify both sketches and code iteratively. As P10 stated, "[To change the code] I would have gone and like updated my drawing." Similarly, P11 noted the importance of "misunderstandings from the schematic and the code [being] easily fixed with some kind of back and forth interaction." Observational data verifies these findings, with 47.4\% of participants revisiting their sketches at least once during the coding process, with an average of three references per participant. This suggests that developers rely on visual information to complement their cognitive models throughout the development process.

\begin{mainbox}{Finding 5 - Code Generation Properties} 
    \textbf{Developers need visual code assistants with clear guidelines and explainability qualities for managing users' expectations.} They often revisit their sketches, indicating the need for dynamic, interactive features to enhance code generation and programming iterations.
\end{mainbox}

Looking forward, participants expressed they would like the integration of sketching capabilities within the development environment. As P3 stated, ``\textit{[would help] even if you are in the code interface, and I can draw directly there}''. A significant proportion of participants (21\%) utilized ChatGPT during coding sessions, averaging five prompts per user. This suggests a demand for interactive tools that can assist in code refinement. As P11 noted, ``\textit{[would be useful] if you had some way to further prompt it into fixing the code itself instead of going manually yourself and trying to fix everything.}''

We found high agreement about possible use cases for visual code assistants, with participants highlighting their utility in three key areas: education, prototyping, and meetings or brainstorming. Six participants emphasized the value of visual code assistants in educational settings. 

P10, for instance, highlighted their potential to aid students in learning code and design: ``\textit{Especially for students learning code and design of code, I think it would be really helpful.}''. 
Seven participants found visual code assistants beneficial for prototyping. P6 mentioned their usefulness for creating high-level implementations or boilerplate code: ``\textit{For high-level implementations like building a boilerplate or skeleton of the tasks, this could prove useful.}'' 
Four participants saw value in using visual code assistants for meetings and idea sharing. P12 envisioned using the tool to assign tasks to interns with a clear structural basis: ``\textit{I can just draw a sketch and give them a basic structure. So they can follow this for an earlier experiment.}'' P17 saw its potential in cross-departmental settings, where visual aids could assist non-technical individuals: ``\textit{I could come up with this maybe during the meeting, especially the cross-department working environment with some non-technical people.}''

These insights reveal broad agreement on the practical applications of visual code assistants, highlighting their potential in education, prototyping, and collaborative settings.

\begin{mainbox}{Finding 6 - Future Use Cases}
Participants identified many applications for visual code assistants, including \textbf{education, prototyping, and collaboration.} Future visual assistants should aim to support iterative refinement of code based on user feedback.
\end{mainbox}

\section{Threats to Validity}

\noindent
\textbf{Construct validity.} A number of metrics were introduced as useful, but imperfect proxies of real-world properties. We used years of programming experience as a proxy for domain knowledge as it is a common metric and all our participants had worked on DS tasks. However, individuals with fewer years of programming experience may possess a greater depth of knowledge in DS. We assessed sketching proficiency based on the recency of the last sketch. This assumption may not hold, but is substantially more cost-effective than formally evaluating drawing proficiency. We also used the time spent sketching as a proxy for sketch quality, assuming that more time leads to more detailed and better sketches, which was true on average but not for all participants. Finally, we developed an accuracy metric to evaluate the quality of the generated code. This metric might be biased or misrepresent the true quality if it includes too much or too little detail. We leave more advanced metrics for future work.

\textbf{External validity.} 
The tasks used in this study were selected for their simplicity to ensure that they could be completed within the 60-minute timeframe. As a result, they may not fully capture the complexity of real-world scenarios typically encountered by Data Scientists. Given that the participants required ca. 20 minutes to complete the tasks on average, we believe that the tasks are nevertheless challenging enough to identify real effects. We made efforts to include a diverse group of participants with varying levels of expertise in DS, including students, researchers, and professional developers. Still, the sample may not fully represent the target population due to selection bias and the small sample size (n=19). This sample size proved sufficient to provide rich qualitative and some significant quantitative findings, but future work may extend this investigation to larger groups to identify more subtle effects. 

\textbf{Internal Validity.}
Given the model's extensive internet-based training, it may have encountered very similar tasks and memorized solutions. However, our data contradict this hypothesis. The LLM demonstrated a clear adherence to participant sketches, even when these diverged from common practices. For example, while an 80/20 train-test split is prevalent, participant P4 opted for a 75/25 split, which the LLM accurately implemented. Similarly, participant P13 incorrectly labeled the target variable, a mistake the LLM replicated. These observations suggest that the LLM was not merely relying on memorized solutions but actively processing and interpreting the provided sketches. We also note that the LLM knowing common DS pipelines does not negate findings on its usefulness for other DS tasks, given that such tasks typically have many elements in common.

\section{Conclusion}
We present the first study of developers using a tool that generates DS notebooks directly from whiteboard sketches. Our work provides key findings on the prevalence \& characteristics of sketching: the majority of developers sketch regularly and mix textual annotations with iconography; the performance of visual LLMs on this task: as judged against a set of criteria, common vLLMs generate notebooks with more than 70\% of the structural elements correct, but implementation details mostly need correcting; the benefits of sketching: even a quick sketch can reduce the coding effort in terms of lines of code written by around half and longer sketch times yield more accurate notebooks. Interviews with participants from both industry and academia offer rich insights into desiderata of future visual coding assistants.

\section*{Acknowledgment}
\noindent
This work was supported by by  Fundacão para a Ciência e Tecnologia (FCT), through the project with reference DOI:10.54499/UIDB/50021/2020 and the scholarship with reference PRT/BD/152343/2021.

\bibliographystyle{IEEEtran}
\bibliography{refs}

\begin{thebibliography}{10}
\providecommand{\url}[1]{#1}
\csname url@samestyle\endcsname
\providecommand{\newblock}{\relax}
\providecommand{\bibinfo}[2]{#2}
\providecommand{\BIBentrySTDinterwordspacing}{\spaceskip=0pt\relax}
\providecommand{\BIBentryALTinterwordstretchfactor}{4}
\providecommand{\BIBentryALTinterwordspacing}{\spaceskip=\fontdimen2\font plus
\BIBentryALTinterwordstretchfactor\fontdimen3\font minus \fontdimen4\font\relax}
\providecommand{\BIBforeignlanguage}[2]{{%
\expandafter\ifx\csname l@#1\endcsname\relax
\typeout{** WARNING: IEEEtran.bst: No hyphenation pattern has been}%
\typeout{** loaded for the language `#1'. Using the pattern for}%
\typeout{** the default language instead.}%
\else
\language=\csname l@#1\endcsname
\fi
#2}}
\providecommand{\BIBdecl}{\relax}
\BIBdecl

\bibitem{williams2000collaborative}
L.~Williams and R.~R. Kessler, ``The collaborative software process,'' in \emph{International Conference on Software Engineering 2000}, 2000.

\bibitem{software_mangano_2010}
\BIBentryALTinterwordspacing
N.~Mangano, A.~Baker, M.~Dempsey, E.~Navarro, and A.~van~der Hoek, ``Software design sketching with calico,'' in \emph{Proceedings of the 25th IEEE/ACM International Conference on Automated Software Engineering}, ser. ASE '10.\hskip 1em plus 0.5em minus 0.4em\relax New York, NY, USA: Association for Computing Machinery, 2010, p. 23–32. [Online]. Available: \url{https://doi.org/10.1145/1858996.1859003}
\BIBentrySTDinterwordspacing

\bibitem{go_cherubini_2007}
\BIBentryALTinterwordspacing
M.~Cherubini, G.~Venolia, R.~DeLine, and A.~J. Ko, ``Let's go to the whiteboard: how and why software developers use drawings,'' in \emph{Proceedings of the SIGCHI Conference on Human Factors in Computing Systems}, ser. CHI '07.\hskip 1em plus 0.5em minus 0.4em\relax New York, NY, USA: Association for Computing Machinery, 2007, p. 557–566. [Online]. Available: \url{https://doi.org/10.1145/1240624.1240714}
\BIBentrySTDinterwordspacing

\bibitem{visual_walny_2011}
J.~Walny, S.~Carpendale, N.~Henry~Riche, G.~Venolia, and P.~Fawcett, ``Visual thinking in action: Visualizations as used on whiteboards,'' \emph{IEEE Transactions on Visualization and Computer Graphics}, vol.~17, no.~12, pp. 2508--2517, 2011.

\bibitem{sketches_baltes_2017}
\BIBentryALTinterwordspacing
S.~Baltes and S.~Diehl, ``Sketches and diagrams in practice,'' in \emph{Proceedings of the 22nd ACM SIGSOFT International Symposium on Foundations of Software Engineering}, ser. FSE 2014.\hskip 1em plus 0.5em minus 0.4em\relax New York, NY, USA: Association for Computing Machinery, 2014, p. 530–541. [Online]. Available: \url{https://doi.org/10.1145/2635868.2635891}
\BIBentrySTDinterwordspacing

\bibitem{maintaining_latoza_2006}
\BIBentryALTinterwordspacing
T.~D. LaToza, G.~Venolia, and R.~DeLine, ``Maintaining mental models: a study of developer work habits,'' in \emph{Proceedings of the 28th International Conference on Software Engineering}, ser. ICSE '06.\hskip 1em plus 0.5em minus 0.4em\relax New York, NY, USA: Association for Computing Machinery, 2006, p. 492–501. [Online]. Available: \url{https://doi.org/10.1145/1134285.1134355}
\BIBentrySTDinterwordspacing

\bibitem{geminiteam2024gemini}
G.~Team \emph{et~al.}, ``Gemini: A family of highly capable multimodal models,'' 2024.

\bibitem{gpt4_openai_2023}
OpenAI, ``Gpt-4 technical report,'' 2023.

\bibitem{chen2021evaluating}
M.~Chen, J.~Tworek, H.~Jun, Q.~Yuan, H.~P. de~Oliveira~Pinto, J.~Kaplan, H.~Edwards, Y.~Burda, N.~Joseph, G.~Brockman, A.~Ray, R.~Puri, G.~Krueger, M.~Petrov, H.~Khlaaf, G.~Sastry, P.~Mishkin, B.~Chan, S.~Gray, N.~Ryder, M.~Pavlov, A.~Power, L.~Kaiser, M.~Bavarian, C.~Winter, P.~Tillet, F.~P. Such, D.~Cummings, M.~Plappert, F.~Chantzis, E.~Barnes, A.~Herbert-Voss, W.~H. Guss, A.~Nichol, A.~Paino, N.~Tezak, J.~Tang, I.~Babuschkin, S.~Balaji, S.~Jain, W.~Saunders, C.~Hesse, A.~N. Carr, J.~Leike, J.~Achiam, V.~Misra, E.~Morikawa, A.~Radford, M.~Knight, M.~Brundage, M.~Murati, K.~Mayer, P.~Welinder, B.~McGrew, D.~Amodei, S.~McCandlish, I.~Sutskever, and W.~Zaremba, ``Evaluating large language models trained on code,'' 2021.

\bibitem{guo2024deepseekcoderlargelanguagemodel}
\BIBentryALTinterwordspacing
D.~Guo, Q.~Zhu, D.~Yang, Z.~Xie, K.~Dong, W.~Zhang, G.~Chen, X.~Bi, Y.~Wu, Y.~K. Li, F.~Luo, Y.~Xiong, and W.~Liang, ``Deepseek-coder: When the large language model meets programming -- the rise of code intelligence,'' 2024. [Online]. Available: \url{https://arxiv.org/abs/2401.14196}
\BIBentrySTDinterwordspacing

\bibitem{roziere2024codellama}
\BIBentryALTinterwordspacing
B.~Rozière, J.~Gehring, F.~Gloeckle, S.~Sootla, I.~Gat, X.~E. Tan, Y.~Adi, J.~Liu, R.~Sauvestre, T.~Remez, J.~Rapin, A.~Kozhevnikov, I.~Evtimov, J.~Bitton, M.~Bhatt, C.~C. Ferrer, A.~Grattafiori, W.~Xiong, A.~Défossez, J.~Copet, F.~Azhar, H.~Touvron, L.~Martin, N.~Usunier, T.~Scialom, and G.~Synnaeve, ``Code llama: Open foundation models for code,'' 2024. [Online]. Available: \url{https://arxiv.org/abs/2308.12950}
\BIBentrySTDinterwordspacing

\bibitem{si2024automatingfrontend}
\BIBentryALTinterwordspacing
C.~Si, Y.~Zhang, Z.~Yang, R.~Liu, and D.~Yang, ``Design2code: How far are we from automating front-end engineering?'' 2024. [Online]. Available: \url{https://arxiv.org/abs/2403.03163}
\BIBentrySTDinterwordspacing

\bibitem{Krenn_2023}
\BIBentryALTinterwordspacing
M.~Krenn, L.~Buffoni, B.~Coutinho, S.~Eppel, J.~G. Foster, A.~Gritsevskiy, H.~Lee, Y.~Lu, J.~P. Moutinho, N.~Sanjabi, R.~Sonthalia, N.~M. Tran, F.~Valente, Y.~Xie, R.~Yu, and M.~Kopp, ``Forecasting the future of artificial intelligence with machine learning-based link prediction in an exponentially growing knowledge network,'' \emph{Nature Machine Intelligence}, vol.~5, no.~11, p. 1326–1335, Oct. 2023. [Online]. Available: \url{http://dx.doi.org/10.1038/s42256-023-00735-0}
\BIBentrySTDinterwordspacing

\bibitem{Ericsson1993}
\BIBentryALTinterwordspacing
K.~A. Ericsson and H.~A. Simon, \emph{Protocol Analysis: Verbal Reports as Data}.\hskip 1em plus 0.5em minus 0.4em\relax The MIT Press, Apr. 1993. [Online]. Available: \url{http://dx.doi.org/10.7551/mitpress/5657.001.0001}
\BIBentrySTDinterwordspacing

\bibitem{follow_walny_2011}
J.~Walny, J.~Haber, M.~Dörk, J.~Sillito, and S.~Carpendale, ``Follow that sketch: Lifecycles of diagrams and sketches in software development,'' in \emph{2011 6th International Workshop on Visualizing Software for Understanding and Analysis (VISSOFT)}, 2011, pp. 1--8.

\bibitem{linking_baltes_2017}
\BIBentryALTinterwordspacing
S.~Baltes, P.~Schmitz, and S.~Diehl, ``Linking sketches and diagrams to source code artifacts,'' in \emph{Proceedings of the 22nd ACM SIGSOFT International Symposium on Foundations of Software Engineering}, ser. FSE 2014.\hskip 1em plus 0.5em minus 0.4em\relax New York, NY, USA: Association for Computing Machinery, 2014, p. 743–746. [Online]. Available: \url{https://doi.org/10.1145/2635868.2661672}
\BIBentrySTDinterwordspacing

\bibitem{roundtrip_baltes_2017}
\BIBentryALTinterwordspacing
S.~Baltes, F.~Hollerich, and S.~Diehl, ``{ Round-Trip Sketches: Supporting the Lifecycle of Software Development Sketches from Analog to Digital and Back },'' in \emph{2017 IEEE Working Conference on Software Visualization (VISSOFT)}.\hskip 1em plus 0.5em minus 0.4em\relax Los Alamitos, CA, USA: IEEE Computer Society, Sep. 2017, pp. 94--98. [Online]. Available: \url{https://doi.ieeecomputersociety.org/10.1109/VISSOFT.2017.24}
\BIBentrySTDinterwordspacing

\bibitem{go_almeida_2022}
\BIBentryALTinterwordspacing
E.~Santana~de Almeida, I.~Ahmed, and A.~van~der Hoek, ``{ Let's Go to the Whiteboard (Again): Perceptions From Software Architects on Whiteboard Architecture Meetings },'' \emph{IEEE Transactions on Software Engineering}, vol.~49, no.~10, pp. 4773--4795, Oct. 2023. [Online]. Available: \url{https://doi.ieeecomputersociety.org/10.1109/TSE.2023.3314410}
\BIBentrySTDinterwordspacing

\bibitem{beltramelli2017pix2code}
\BIBentryALTinterwordspacing
T.~Beltramelli, ``pix2code: Generating code from a graphical user interface screenshot,'' in \emph{Proceedings of the ACM SIGCHI Symposium on Engineering Interactive Computing Systems}, ser. EICS '18.\hskip 1em plus 0.5em minus 0.4em\relax New York, NY, USA: Association for Computing Machinery, 2018. [Online]. Available: \url{https://doi.org/10.1145/3220134.3220135}
\BIBentrySTDinterwordspacing

\bibitem{learning_ellis_2018}
K.~Ellis, D.~Ritchie, A.~Solar-Lezama, and J.~B. Tenenbaum, ``Learning to infer graphics programs from hand-drawn images,'' in \emph{Proceedings of the 32nd International Conference on Neural Information Processing Systems}, ser. NIPS'18.\hskip 1em plus 0.5em minus 0.4em\relax Red Hook, NY, USA: Curran Associates Inc., 2018, p. 6062–6071.

\bibitem{psdoodle_mohian_2022}
\BIBentryALTinterwordspacing
S.~Mohian and C.~Csallner, ``Psdoodle: fast app screen search via partial screen doodle,'' in \emph{Proceedings of the 9th IEEE/ACM International Conference on Mobile Software Engineering and Systems}, ser. MOBILESoft '22.\hskip 1em plus 0.5em minus 0.4em\relax New York, NY, USA: Association for Computing Machinery, 2022, p. 89–99. [Online]. Available: \url{https://doi.org/10.1145/3524613.3527816}
\BIBentrySTDinterwordspacing

\bibitem{dlpaper2code_sethi_2017}
\BIBentryALTinterwordspacing
A.~Sethi, A.~Sankaran, N.~Panwar, S.~Khare, and S.~Mani, ``Dlpaper2code: Auto-generation of code from deep learning research papers,'' \emph{Proceedings of the AAAI Conference on Artificial Intelligence}, vol.~32, no.~1, Apr. 2018. [Online]. Available: \url{http://dx.doi.org/10.1609/aaai.v32i1.12326}
\BIBentrySTDinterwordspacing

\bibitem{mathvista24}
\BIBentryALTinterwordspacing
P.~Lu, H.~Bansal, T.~Xia, J.~Liu, C.~Li, H.~Hajishirzi, H.~Cheng, K.-W. Chang, M.~Galley, and J.~Gao, ``Mathvista: Evaluating mathematical reasoning of foundation models in visual contexts,'' 2024. [Online]. Available: \url{https://arxiv.org/abs/2310.02255}
\BIBentrySTDinterwordspacing

\bibitem{yue2024mmmu}
\BIBentryALTinterwordspacing
X.~Yue, Y.~Ni, K.~Zhang, T.~Zheng, R.~Liu, G.~Zhang, S.~Stevens, D.~Jiang, W.~Ren, Y.~Sun, C.~Wei, B.~Yu, R.~Yuan, R.~Sun, M.~Yin, B.~Zheng, Z.~Yang, Y.~Liu, W.~Huang, H.~Sun, Y.~Su, and W.~Chen, ``Mmmu: A massive multi-discipline multimodal understanding and reasoning benchmark for expert agi,'' 2024. [Online]. Available: \url{https://arxiv.org/abs/2311.16502}
\BIBentrySTDinterwordspacing

\bibitem{pădurean2024}
\BIBentryALTinterwordspacing
V.-A. Pădurean and A.~Singla, ``Benchmarking generative models on computational thinking tests in elementary visual programming,'' 2024. [Online]. Available: \url{https://arxiv.org/abs/2406.09891}
\BIBentrySTDinterwordspacing

\bibitem{wu2024plot2code}
\BIBentryALTinterwordspacing
C.~Wu, Y.~Ge, Q.~Guo, J.~Wang, Z.~Liang, Z.~Lu, Y.~Shan, and P.~Luo, ``Plot2code: A comprehensive benchmark for evaluating multi-modal large language models in code generation from scientific plots,'' 2024. [Online]. Available: \url{https://arxiv.org/abs/2405.07990}
\BIBentrySTDinterwordspacing

\bibitem{li2024mmcode}
\BIBentryALTinterwordspacing
K.~Li, Y.~Tian, Q.~Hu, Z.~Luo, and J.~Ma, ``Mmcode: Evaluating multi-modal code large language models with visually rich programming problems,'' 2024. [Online]. Available: \url{https://arxiv.org/abs/2404.09486}
\BIBentrySTDinterwordspacing

\bibitem{visual_damorim_2020}
M.~d’Amorim, R.~Abreu, and C.~Mello, ``Visual sketching: From image sketches to code,'' in \emph{2020 IEEE/ACM 42nd International Conference on Software Engineering: New Ideas and Emerging Results (ICSE-NIER)}, 2020, pp. 101--104.

\bibitem{yujie2024wildvisionarena}
\BIBentryALTinterwordspacing
Y.~Lu, D.~Jiang, W.~Chen, W.~Wang, Y.~Choi, and B.~Y. Lin, ``Wildvision arena: Benchmarking multimodal llms in the wild,'' February 2024. [Online]. Available: \url{https://huggingface.co/spaces/WildVision/vision-arena/}
\BIBentrySTDinterwordspacing

\bibitem{2023opencompass}
O.~Contributors, ``Opencompass: A universal evaluation platform for foundation models,'' \url{https://github.com/open-compass/opencompass}, 2023.

\bibitem{easterbrook2008selecting}
S.~Easterbrook, J.~Singer, M.-A. Storey, and D.~Damian, ``Selecting empirical methods for software engineering research,'' \emph{Guide to advanced empirical software engineering}, pp. 285--311, 2008.

\bibitem{singer2008software}
J.~Singer, S.~E. Sim, and T.~C. Lethbridge, ``Software engineering data collection for field studies,'' \emph{Guide to advanced empirical software engineering}, pp. 9--34, 2008.

\bibitem{zheng2023judgingllm}
\BIBentryALTinterwordspacing
L.~Zheng, W.-L. Chiang, Y.~Sheng, S.~Zhuang, Z.~Wu, Y.~Zhuang, Z.~Lin, Z.~Li, D.~Li, E.~P. Xing, H.~Zhang, J.~E. Gonzalez, and I.~Stoica, ``Judging llm-as-a-judge with mt-bench and chatbot arena,'' 2023. [Online]. Available: \url{https://arxiv.org/abs/2306.05685}
\BIBentrySTDinterwordspacing

\bibitem{chiang2024chatbot}
W.-L. Chiang, L.~Zheng, Y.~Sheng, A.~N. Angelopoulos, T.~Li, D.~Li, H.~Zhang, B.~Zhu, M.~Jordan, J.~E. Gonzalez, and I.~Stoica, ``Chatbot arena: An open platform for evaluating llms by human preference,'' 2024.

\bibitem{huang2024limitationsfinetunedjudgemodels}
\BIBentryALTinterwordspacing
H.~Huang, Y.~Qu, H.~Zhou, J.~Liu, M.~Yang, B.~Xu, and T.~Zhao, ``On the limitations of fine-tuned judge models for llm evaluation,'' 2024. [Online]. Available: \url{https://arxiv.org/abs/2403.02839}
\BIBentrySTDinterwordspacing

\end{thebibliography}

\end{document}